\documentclass[pra,twocolumn,showpacs]{revtex4}%
\usepackage{graphicx}
\usepackage{graphicx}
\usepackage{amsmath}
\usepackage{amsfonts}
\usepackage{amssymb}
\usepackage{hyperref}%
\providecommand{\U}[1]{\protect\rule{.1in}{.1in}}
\ifx\pdftexversion\undefined \else \fi
\expandafter\ifx\csname package@font\endcsname\relax\else
\expandafter\expandafter \expandafter \expandafter
\expandafter{\csname package@font\endcsname} \fi
\begin{document}
\title{A high-flux 2D MOT source for cold lithium atoms}
\date{\today}
\author{T.G. Tiecke}
\author{S.D. Gensemer}
\altaffiliation{Present address: Ethel Walker School, 230 Bushy Hill Rd, Simsbury, CT 06070,
United States}

\author{A. Ludewig}
\author{J.T.M. Walraven}
\affiliation{Van der Waals-Zeeman Institute of the University of Amsterdam,
Valckenierstraat 65, 1018 XE, The Netherlands}

\begin{abstract}
We demonstrate a novel 2D MOT beam source for cold $^{6}\mathrm{Li}$ atoms.
The source is side-loaded from an oven operated at temperatures in the range
$600\lesssim T\lesssim700~\mathrm{K.}$ The performance is analyzed by loading
the atoms into a 3D MOT located $220$~\textrm{mm} downstream from the source.
The maximum recapture rate of $\sim10^{9}~\mathrm{s^{-1}}$ is obtained for
$T\approx700~\mathrm{K}$ and results in a total of up to $10^{10}$ trapped
atoms. The recaptured fraction is estimated to be $30\pm10\%$ and limited by
beam divergence. The most-probable velocity in the beam $(\alpha_{z})$ is
varied from $18$ to $70$~\textrm{m/s} by\textrm{ }increasing the intensity of
a push beam. The source is quite monochromatic with a full-width at half
maximum velocity spread of $11$~\textrm{m/s} at $\alpha_{z}=36~\mathrm{m/s}$,
demonstrating that side-loading completely eliminates beam contamination by
hot vapor from the oven. We identify depletion of the low-velocity tail of the
oven flux as the limiting loss mechanism$.$ Our approach is suitable for other
atomic species.

\end{abstract}

\maketitle

\section{Introduction}

Since the first demonstration of a laser-cooled atomic beam by Phillips and
Metcalf \cite{phillips82} the development and improvement of cold atom sources
has evolved into an essential activity in atomic physics laboratories. In
particular sources for cold Rb, K and Cs received a lot of attention and
became compact and familiar standard devices \cite{metcalf99}. However, for
most other atomic and all molecular species the situation is less favorable
and considerable time as well as resources remain necessary for the
development of a source. Aside from optical cooling schemes many other cooling
principles have been explored, we mention cryogenic cooling by surfaces
\cite{silvera80} or buffer gas \cite{doyle95}, filtering by magnetic
\cite{hulet99,nikitin03} or electric funnels \cite{pinkse04} and Stark
deceleration of molecules \cite{meijer99} as well as Rydberg atoms
\cite{merkt06}. In spite of the success of these sources in specific cases,
optical cooling is the preferred option whenever an appropriate optical
transition is available.

The highest optically cooled atom fluxes to date have been produced from
Zeeman-slowed atomic beams \cite{meschede00,hau05,ketterle05,straten07}.
Zeeman slowers have the additional advantage of a wide applicability.
Unfortunately, their use adds a substantial engineering effort to system
design and construction, in particular if beam-brightening and recycling
principles are involved \cite{hau94,hau05}. The magnetic field inside the
Zeeman slower must be very smooth and satisfy a particular profile in order to
optimize the slowing. In addition, as the acceptance angle is small, the
source oven has to be positioned on the beam axis and operated under high flux
conditions. In typical applications this gives rise to a high background of
hot atoms and results in maintenance because the oven has to be reloaded regularly.

An important simplification of cold atom sources was realized when Monroe
\textit{et.~al.~}\cite{monroe90} demonstrated that in a room-temperature vapor
a fraction of the atoms can be optically captured and cooled into a
magneto-optical trap (MOT) and subsequently loaded into a magnetic trap. The
primary drawback of this vapor-cell MOT (VCMOT) is that the lifetime of the
magnetically trapped atoms is limited by collisions with hot atoms from the
vapor, thus limiting the time available for experiment. One approach to
overcome this limitation is pulsed loading, starting from an alkali getter
dispenser \cite{stamper05} or by ultraviolet light induced desorption
\cite{gozzini93,arlt06}. All other solutions involve a dual chamber
arrangement in which a source chamber, containing some variation of the VCMOT
source, is separated by a differential pumping channel from an
ultra-high-vacuum (UHV) chamber in which the atoms are recaptured in a
secondary MOT in preparation for experiments under UHV conditions .

Three basic types of VCMOT sources are used in the dual MOT configurations. In
the first type a pulsed VCMOT serves to load the recapture MOT by a sequence
of cold atom bunches, transferred with the aid of a push beam \cite{wieman96}.
The second type is known as the LVIS (low-velocity intense source)
\cite{cornell96}. In this case the VCMOT and the push beam are operated
continuously, giving rise to a steady beam of cold atoms in the direction of
the push beam. In the third type the standard three-dimensional (3D) MOT
arrangement in the source chamber is replaced by a two-dimensional (2D) MOT
configuration, with (2D$^{+}$-MOT) or without (2D~MOT) push and cooling beams
along the symmetry axis \cite{dieckmann98,pfau02,inguscio06}. This has the
important advantage that the source MOT can be optimized for capture because,
with confinement in only two directions, the residence time and collisional
losses are intrinsically low.

VCMOT sources work most conveniently for elements like Cs, Rb, and K, having a
vapor pressure of $\sim10^{-7}$ mbar around room temperature \cite{alcock84}.
Elements such as Li, Yb, Cr and the alkaline earths must be loaded from atomic
beams since their vapor pressures are only significant at temperatures far
above the maximum baking temperature of a conventional UHV system
\cite{alcock84,ketterle05,yabuzaki99,pfau05}. In the case of elements which
are chemically reactive with glass, such as Li, a vapor cell is additionally impractical.

In this paper we present a novel 2D~MOT source for cold lithium. It yields a
cold flux comparable to the maximum achieved with lithium Zeeman slowers
\cite{priv08}. Contrary to previously realized 2D~MOT systems our source is
transversely loaded with a beam from an effusive oven, rather than
longitudinally like in beam brighteners or isotropically like in vapor cells.
This demonstrates the possibility to use 2D~MOT sources in applications where
a vapor cell cannot be used and avoids the background of hot atoms in the
beam. An important \textit{a priory} uncertainty of this arrangement is the
risk of depletion of the low-velocity tail of capturable atoms by the onset of
nozzling as occurred in the famous Zacharias fountain experiment
\cite{estermann47,vanier89}. Our work shows that large cold atomic fluxes can
be realized without this depletion becoming inhibitive. Recently this was also
demonstrated with a Li oven loaded 3D~MOT \cite{madison09}. Another novelty of
our source is the application of the 2D~MOT concept to a light atom like
lithium. Magneto-optical trapping of light species requires a high gradient
for efficient capture. As this also compresses the cold atoms into a cloud of
small volume, in particular in the 3D configuration trap losses are
substantial even for small atom numbers. We demonstrate that in our dual MOT
arrangement, the 2D~MOT can be optimized for capture with a large gradient and
without considerable losses, whereas the 3D recapture MOT can be optimized
with a different gradient for maximum total atom number.

In the following sections we describe our experimental apparatus (section
\ref{section:Experimental}) and our results (section
\ref{section:ExperimentalResults}). In section \ref{section:SourceModel} we
present a simple model for the loading of the 2D MOT. The performance of our
system and loss mechanisms are discussed in section \ref{section:Discussion}
and in section \ref{section:Conclusion} we summarize our findings and comment
on the suitability of our approach for other atomic species.

\section{Experimental\label{section:Experimental}}

\subsection{Vacuum system}

\begin{figure}
[ptb]
\begin{center}
\includegraphics[
natheight=5.441600in,
natwidth=3.681000in,
height=11.4652cm,
width=7.7704cm
]%
{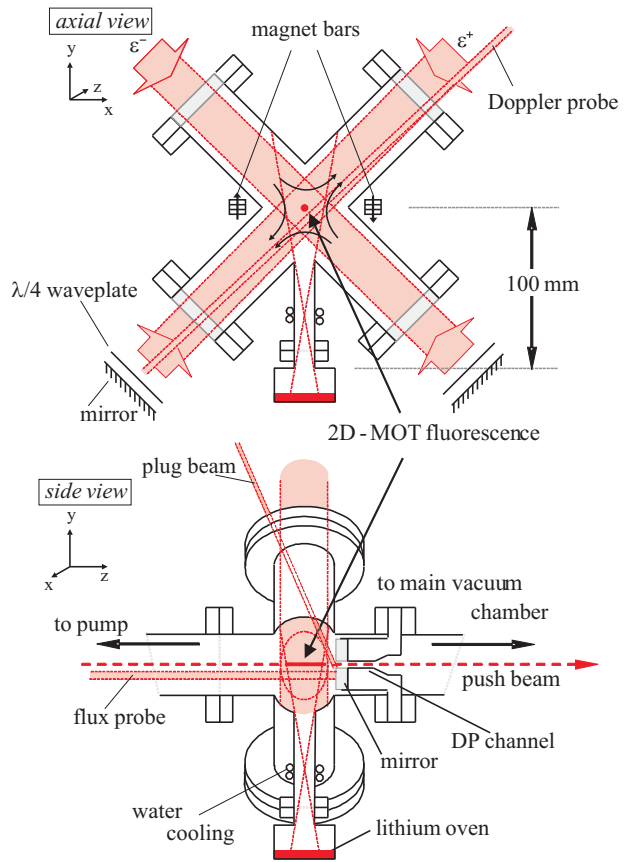}%
\caption{Schematic drawing of the 2D-MOT system. The oven tube is welded into
the center of a six way cross as described in the text. \textit{Upper
drawing}: vertical cross section through the oven viewing along the beam axis;
\textit{lower drawing}: vertical cross section through the oven and through
the DP-channel viewing the beam from the side. The Doppler probe is under 50
degrees with the vertical ($y$) axis and is used to calibrate the oven
temperature. The flux probe is used to measure the hot flux emitted by the
oven; a gold-plated mirror is included for this purpose. The plug beam is used
to interupt the atomic beam for time-of-flight measurements. The
two-dimensional quadrupole field required for the 2D~MOT is provided by two
permanent-magnet bars.}%
\label{fig:vacuum}%
\end{center}
\end{figure}

The experimental setup of the lithium 2D~MOT source is sketched in
Fig.\thinspace\ref{fig:vacuum}. The vacuum system consists of a stainless
steel six-way cross of $40\mathrm{~mm}$ tubing of which two CF40 ports define
the horizontal symmetry axis of the source. The other four CF40 ports are
configured under $45^{\circ}$ and sealed with standard vacuum windows
providing the optical access for the retroreflected 2D~MOT beams with a waist
($1/e^{2}$ radius) $w=9\mathrm{~mm}$. A lithium oven is mounted with a CF16
flange onto the bottom of a water-cooled tube with inner radius
$a=8\mathrm{~mm}$ and connected along the vertical axis into the center of the
cross. The source is connected horizontally onto the main UHV chamber,
separated by a gate valve. Between the main vacuum and the source a
$23\mathrm{~mm}$ long differential pumping (DP) channel of $2\mathrm{~mm}$
diameter can maintain a maximum pressure ratio of $10^{-3}$ between the main
UHV chamber and the source. There is no direct line of sight from the oven to
the main UHV chamber nor to the windows. When the oven is operated and the
2D~MOT lasers are off, no lithium was detected in the main UHV chamber. Also
no measurable gas load is observed on the main vacuum while the source is
operated.%

\subsection{Lithium oven}

The oven consists of a stainless steel lithium reservoir, $25\mathrm{~mm}$
high and $50\mathrm{~mm}$ in diameter, attached to a CF16 flange by a
$15\mathrm{~mm}$ long tube of $16\mathrm{~mm}$ inner diameter. The oven is
embedded in a simple heat shield of glass wool and aluminium foil and is
connected to the vacuum system using a nickel gasket. The reservoir was loaded
with $\sim6$ g of $^{6}$Li and $\sim2$ g of $^{7}$Li under an inert gas
(argon) atmosphere. As commercial lithium contains a large fraction of LiH it
has to be degassed by dissociating the hydride. For this purpose we baked the
oven under vacuum in a separate setup for two hours at a temperature of
$\sim943~\mathrm{K}$. Some $25\%$ of the lithium was lost in this process. To
protect the employed turbopump from alkali contamination a liquid nitrogen
cold trap was used in this procedure.

Under typical conditions the oven is operated at $T=623(12)\mathrm{~K}$ ($350$
C), well above the melting point of lithium at $454\mathrm{~K}$. All data
presented in this paper, except those presented in Fig.\thinspace
\ref{fig:loadingvstemp}, were obtained at this temperature. The oven
temperature is calibrated by Doppler thermometry of the emerging Li flux using
a probe beam under $50^{\circ}$ with the vertical axis (see Fig.\thinspace
\ref{fig:vacuum}). Temperature stabilization is done with a thermocouple
reference. Starting from room temperature the oven reaches the regulated value
of $623\mathrm{~K}$ in $\sim15$ minutes. The $^{6}$Li abundance was measured
to be $a_{6}=0.74(5)$ using absorption spectroscopy on the $^{6}$Li $D_{1}$
$(^{2}S_{1/2}\rightarrow^{2}P_{1/2})$ line and the $^{7}$Li $D_{2}$
$(^{2}S_{1/2}\rightarrow^{2}P_{3/2})$ line.

\subsection{The 2D~MOT configuration}

As sketched in Fig.\thinspace\ref{fig:vacuum} the 2D~MOT consists of a 2D
quadrupole magnetic field in combination with two orthogonal pairs of
retroreflected laser beams of opposite circular polarization, at a power of up
to $50$ mW per beam in a waist of $9\mathrm{~mm}$ and red-detuned with respect
to the optical resonance near $671~\mathrm{nm}$. Like in a standard 3D~MOT
\cite{metcalf99}, a cold atom moving in the crossed laser field is optically
pumped to a state for which the Zeeman shift places it closer to resonance
with a laser opposing the motion of the atom. Thus the atoms are trapped and
cooled in the radial direction and collect along the symmetry axis of the 2D
quadrupole field but are free to move in the axial direction. As a result only
atoms with a sufficiently low axial velocity can be radially trapped; atoms
with a residence time of less then $0.5$~ms in the optical trapping region
leave the 2D~MOT before they are significantly cooled. Only the radially
cooled atoms give rise to a sufficiently collimated beam to pass through the
DP-channel and be recaptured by a 3D MOT in the middle of the UHV chamber.

For best performance the atoms are accelerated out of the source by a push
beam, aligned along the symmetry axis and with a waist of $1.2$~mm passing
through the DP-channel. The detuning and intensity of the push laser determine
the velocity of the atoms emerging from the source. This velocity is chosen
below the capture limit of the recapture MOT but is sufficiently fast to
assure that the atoms do not fall below the recapture region as a result of
gravity. For this reason the push beam is essential for horizontal
configurations but optional in vertical arrangements. In all arrangements the
push beam acceleration increases the output flux because it reduces the
residence time in the 2D~MOT and therefore background-induced losses. In the
literature on the 2D$^{+}$ MOT \cite{dieckmann98,pfau02,inguscio06} and the
LVIS \cite{cornell96} control over the axial velocity is reported by using a
pair of counter-propagating axial cooling beams over the entire trap but this
method is not employed here.

The magnetic quadrupole field is provided by two sets of Nd$_{2}$Fe$_{14}$B
magnets (Eclipse magnets N750-RB) with a measured magnetization of
$8.8(1)\times10^{5}~\mathrm{A\,m}^{-1}$. Each set consists of two stacks of
three $25\times10\times3~\mathrm{mm}$ magnet bars separated by $12~\mathrm{mm}%
$ to make an effective dipole bar of $62\mathrm{~mm}$ total length. The
optimum position of the centers of the dipole bars was experimentally found to
be $x=\pm42\mathrm{~mm}$ from the symmetry axis in the horizontal plane as
sketched in Fig.\thinspace\ref{fig:vacuum}. For this distance we calculate a
field gradient of $0.50~\mathrm{T/m}$, constant within $2\%$ along the 2D~MOT
symmetry axis over a total length of $20\mathrm{~mm}$. The use of permanent
magnets simplifies the application of the high field gradients needed for
light species. It combines a simple construction with convenient alignment and
occupies much less space than the more traditional racetrack coils. The
quadrupole field falls off over short distances along the symmetry axis. At
the position of the recapture MOT, $23$ cm downstream from the center of the
2D~MOT, only a small gradient of $210~\mathrm{\mu T/m}$ remains.%
\begin{figure}
[ptb]
\begin{center}
\includegraphics[
trim=0.596258in 0.245717in 0.353099in 0.250812in,
natheight=2.315900in,
natwidth=2.804600in,
height=6.9751cm,
width=7.108cm
]%
{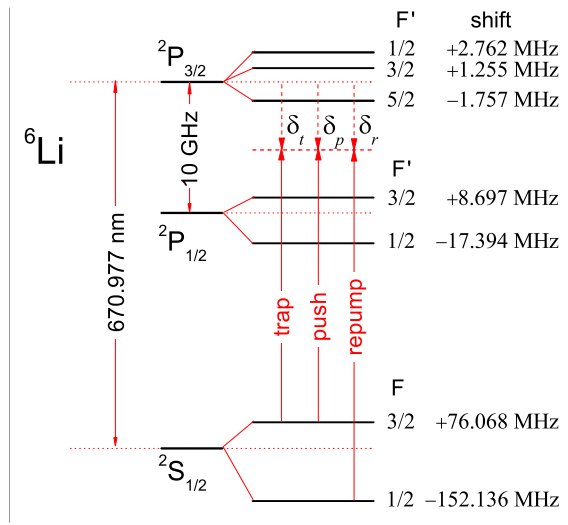}%
\caption{Level structure of $^{6}$Li. Note that the hyperfine splitting of the
$^{2}P_{3/2}$ levels is smaller than the natural linewidth $\Gamma/2\pi
=5.9$~MHz of the $D_{2}$ $\left(  ^{2}S_{1/2}\rightarrow^{2}P_{3/2}\right)  $
transition.}%
\label{fig:LiLevels}%
\end{center}
\end{figure}

\subsection{Hyperfine structure of $^{6}\mathrm{Li}$ levels}

Laser cooling of $^{6}\mathrm{Li}\,$ differs from the familiar case of\textit{
}$^{87}$Rb, in which a spectrally-well-resolved cycling transition on the
$D_{2}$ line can be strongly driven to cool and trap the atoms while a weak
repumping beam is sufficient to compensate for parasitic leakage to the dark
state manifold. In the case of $^{6}\mathrm{Li}$ the hyperfine splitting of
the $2^{2}P_{3/2}$ excited state is of the order of the natural linewidth,
$\Gamma/2\pi=5.9\mathrm{~MHz}$ and all $D_{2}$ transitions from the $F=3/2$
manifold, $|2^{2}S_{1/2};F=3/2\rangle\rightarrow|2^{2}P_{3/2};F^{\prime
}=1/2,3/2,5/2\rangle$ are excited simultaneously (see Fig.\thinspace
\ref{fig:LiLevels}) \cite{das07}. Hence, there is no closed transition
suitable for cooling and trapping and strong optical pumping to the
$|2^{2}S_{1/2};F=1/2\rangle$ level cannot be avoided. As a consequence the
`trapping' and `repumping' beams have to be of comparable intensities, which
means that both beams contribute to the cooling and mutually serve for
repumping. Also the detunings will have a strong influence in this respect
\cite{grimm98}. In spite of these differences we stick to the conventional
terminology, referring to the transition $|^{2}S_{1/2};F=3/2\rangle
\rightarrow|^{2}P_{3/2};F^{\prime}=1/2,3/2,5/2\rangle$ as the `trapping'
transition and to $|^{2}S_{1/2};F=1/2\rangle\rightarrow|^{2}P_{3/2};F^{\prime
}=1/2,3/2\rangle$ as the `repumping' transition.

\subsection{Laser system}

A laser system for wavelength $\lambda_{L}=671\mathrm{~nm}$ was developed to
serve the 2D (source) MOT and the 3D (recapture) MOT as well as to provide
laser beams for $^{6}$Li diagnostics. The laser system consists of a single
master oscillator and four injection-locked slave lasers, all operating a
$120\mathrm{~mW}$ Mitshubishi ML101J27 diode heated to $70\mathrm{~C}$. The
master oscillator is a home-built external-cavity diode laser (ECDL)
\cite{haensch95}, frequency stabilized using saturated absorption spectroscopy
in a $^{6}$Li heat pipe \cite{vidal69}. The power from the master laser is
distributed over six beams, which can be independently shifted in frequency
using ISOMET 1205-C acousto-optic modulators (AOM's). Of these six beams four
are amplified by injection-locking of the slave lasers and of these four beams
one pair is used for the retroreflected trapping and repumping beams of the
2D~MOT while the other pair is equally distributed over six beams and
similarly employed for the 3D~MOT. The remaining two frequency-shifted ECDL
beams serve as pushing beam, as probing beam or as plug beam in various
diagnostic applications.

\section{Source model\label{section:SourceModel}}

\subsection{Oven flux}

To establish the principle of our source and to enable comparison with
experiment we present a semi-empirical kinetic model in which the oven is
replaced by an emittance of area $A=\pi a^{2}\approx2~\mathrm{cm}^{2}$ at the
saturated vapor pressure of lithium. Around $T=623~\mathrm{K}$ the saturated
vapor pressure is given by $p_{s}=p_{a}\exp(-L_{0}/k_{B}T)$ where
$p_{a}=1.15(5)\times10^{10}$ Pa and $L_{0}/k_{B}=18474~\mathrm{K}$ is the
latent heat of vaporization \cite{alcock84}. As $p_{s}$ is only accurate to
within $5\%$ we neglect the small dependence on the isotopic composition. The
total atomic flux $\Phi_{tot}$ emitted by the oven may be estimated by the
detailed balance expression for the total flux onto and from the emittance
under thermal equilibrium conditions,
\begin{equation}
\Phi_{tot}=\frac{1}{4}n_{s}\bar{v}\,A, \label{Exp:flux-b}%
\end{equation}
where $n_{s}$ is the atomic density and $\bar{v}=\left[  8k_{B}T/\pi m\right]
^{1/2}$ the mean thermal speed, with $k_{B}$ the Boltzmann constant and $m$
the mass of the Li atoms. For $T=623(12)~\mathrm{K}$ we have $p_{s}%
=1.5_{-0.7}^{+1.1}\times10^{-3}$ Pa, corresponding to a density $n_{s}%
=1.8_{-0.8}^{+1.2}\times10^{17}$~$\mathrm{m}^{-3}$. With these numbers the
total flux from the source is found to be $\Phi_{tot}\approx1.3\times
10^{16}\mathrm{~s}^{-1}\approx1.3\times10^{-10}~\mathrm{kg\,s}^{-1}$. With $8$
gram of Li this corresponds to $\sim17000$ hours running time.

The flux of $^{6}$Li atoms captured by a 2D MOT at a distance of
$L=100\mathrm{~mm}$ above the oven can be written as an integral over the
velocity distribution%
\begin{equation}
\Phi_{c}=a_{6}n_{s}A\int_{0}^{\Omega_{c}}d\Omega\frac{\cos\theta}{4\pi}%
\frac{1}{\mathcal{N}}\int_{0}^{v_{c}}v^{3}e^{-\left(  v/\alpha\right)  ^{2}%
}dv, \label{Exp:flux-a}%
\end{equation}
where $a_{6}=0.74(5)$ is the $^{6}$Li abundance, $\Omega_{c}=A_{c}%
/L^{2}=2\times10^{-2}$ the solid angle of capture (with $A_{c}$ the capture
surface), $d\Omega=2\pi\sin\theta d\theta$ with $\theta$ the emission angle
with respect to the oven axis, $v_{c}$ the capture velocity, $\alpha=\left[
2k_{B}T_{0}/m\right]  ^{1/2}=1.31\times10^{3}~$\textrm{m/s} the most-probable
atomic speed in the oven and $\mathcal{N}=\int v^{2}e^{-\left(  v/\alpha
\right)  ^{2}}dv=\pi^{1/2}\alpha^{3}/4$ the normalization factor of the speed
distribution. Note that by integrating Eq.\thinspace(\ref{Exp:flux-a}) over a
hemisphere we regain Eq.\thinspace(\ref{Exp:flux-b}) in the limit $\left(
a_{6}\rightarrow1;v_{c}\rightarrow\infty\right)  $. Because the solid angle of
capture is small we have $\cos\theta\simeq1$ and the flux $\Phi_{s}$ emitted
by the oven within the solid angle of capture is given by
\begin{equation}
\Phi_{s}\simeq n_{s}\bar{v}\,A\frac{\Omega_{c}}{4\pi}. \label{eq:Phis}%
\end{equation}
For $T=623(12)~\mathrm{K}$ we calculate a total flux density of $\Phi
_{s}/A_{c}=4_{-1.6}^{+3.2}\times10^{13}$ s$^{-1}$cm$^{-2}$ at
$L=100\mathrm{~mm}$ above the oven. Presuming the capture speed to be small,
$v_{c}\ll\alpha$, the captured flux $\Phi_{c}$ may be approximated by
\begin{equation}
\Phi_{c}\simeq\frac{1}{2}a_{6}n_{s}\bar{v}\,A\left(  \frac{v_{c}}{\alpha
}\right)  ^{4}\frac{\Omega_{c}}{4\pi}=\frac{1}{2}a_{6}\left(  \frac{v_{c}%
}{\alpha}\right)  ^{4}\Phi_{s} \label{eq:PhiMax}%
\end{equation}
This expression represents the theoretical maximum flux that can be extracted
from the 2D~MOT source.

\subsection{Capture and cooling}

\label{sect:radialCapt}To discuss the capture and cooling behavior in the
2D~MOT we distinguish two coaxial spatial regions, crossing-over at $r=r_{d}$
defined by $\delta_{Z}(r_{d})+\delta_{L}=0$, \textit{i.e.}~the surface where
the Zeeman shift in the radial gradient of the quadrupole field, $\hbar
\delta_{Z}\left(  r\right)  =\mu_{B}\left(  \partial B/\partial r\right)  r$,
is compensated by the detuning of the laser, $\delta_{L}=\omega_{L}-\omega
_{0}<0$, \textit{i.e.}, to the red side of the cooling transition at angular
frequency $\omega_{0}$ in zero field.

In the outer region $\left(  r>r_{d}\right)  $, the 2D~MOT functions much like
a Zeeman slower, while in the inner region $\left(  r<r_{d}\right)  $ the
motion of the atoms can be described by a damped harmonic oscillator model
\cite{metcalf99}. First we discuss the outer region. An atom with velocity
$\mathbf{v}$ at distance $r$ from the symmetry axis will be at resonance with
the cooling laser if the difference of the Zeeman shift and the laser detuning
equals the Doppler shift,%
\begin{equation}
\delta_{Z}-\delta_{L}=-\mathbf{\mathbf{k}\cdot\mathbf{v}.}
\label{ResCondition}%
\end{equation}
Here $k=|\mathbf{k}|=2\pi/\lambda_{L}$ is the wavevector of the cooling laser.
In view of the angle of $135^{\circ}$ between the directions of the hot
lithium beam and the opposing laser cooling beams the positive Doppler shift
is reduced by a factor $-\mathbf{\cos(\mathbf{k},\mathbf{v})}=\sqrt
{\text{{\small 1/2}}}$ with respect to the fully counter-propagating
configuration. Accordingly, the maximum available slowing distance is larger,
$r_{\max}=\sqrt{2}w$, where $w=9\mathrm{~mm}$ is the waist of the
cooling\ beams. Substituting $r_{\max}$ in the expression for the Zeeman shift
we rewrite Eq.\thinspace(\ref{ResCondition}) in the form of an expression for
the highest atomic speed $v_{\max}$ for which the resonance condition is
satisfied%
\begin{equation}
v_{\max}=\lambda_{L}\frac{\sqrt{2}}{2\pi}\left[  \frac{\mu_{B}}{\hbar}%
\frac{\partial B}{\partial r}r_{\max}-\delta_{L}\right]  . \label{delta}%
\end{equation}
Note that with the left-circular $\left(  \epsilon^{+}\right)  $ and
right-circular $\left(  \epsilon^{-}\right)  $ polarizations of the 2D~MOT
beams as indicated in Fig.\thinspace\ref{fig:vacuum} the atoms are $\sigma
^{+}$ optically pumped into a fully stretched state with the magnetic field
being orthogonal to the propagation direction of the hot flux. In the simplest
1D model for capture process (in which only the trajectory along the symmetry
axis of the oven is considered) $r_{\max}$ represents the capture radius
$\left(  r_{c}\right)  $ and $v_{\max}$ the capture velocity $\left(
v_{c}\right)  $ of the 2D~MOT provided the resonant photon scattering force
$\left(  mdv/dt=\hbar k\Gamma/2\right)  $ is large enough to keep the atom in
resonance with the cooling laser, $\hbar d\delta_{Z}/dt=-\mu_{B}\left(
\partial B/\partial r\right)  v_{\max}$. The resulting condition%
\begin{equation}
v_{\max}\leq\frac{\sqrt{1/8}\left(  \hbar k\right)  ^{2}}{m\mu_{B}\left(
\partial B/\partial r\right)  }\Gamma\label{eq:vmax}%
\end{equation}
is satisfied in our experiment. Combining Eqs.\thinspace(\ref{delta}) and
(\ref{eq:vmax}) we obtain an equation quadratic in $\left(  \partial
B/\partial r\right)  $, which reduces for $\delta_{Z}\gg\delta_{L}$ to%
\begin{equation}
\frac{\partial B}{\partial r}\leq\frac{\left(  \hbar k\right)  ^{3/2}}%
{2\mu_{B}\left(  mr_{\max}\right)  ^{1/2}}\Gamma^{1/2}. \label{eq:-m1/2}%
\end{equation}
This expression shows that the optimal gradient for capture scales like
$m^{-1/2}$, which is important for comparing the performance of the source for
different atomic species. Substituting the optimal gradient into
Eq.\thinspace(\ref{eq:vmax}) we obtain
\begin{equation}
v_{\max}=\left(  a_{\max}r_{\max}\right)  ^{1/2}, \label{eq:vmax2}%
\end{equation}
where $a_{\max}=\hbar k\Gamma/2m$ is the maximum attainable deceleration by
the scattering force.

In spite of the insight it offers the 1D model is far too simple to justify
the use $v_{c}=v_{\max}$ for reliable estimates of the captured flux.
Therefore, we decided to estimate $v_{c}$ experimentally by measuring the
loading rate of the 3D~MOT as a function of the mean velocity in the cold beam
and Eq.\thinspace(\ref{delta}) is only used for scaling between the conditions
of the 3D~MOT and the 2D~MOT. This procedure is discussed in section
\ref{section:Discussion}.

In the inner region $\left(  r<r_{d}\right)  $ of the trap the atomic motion
is described by an overdamped harmonic oscillator model with a spring constant
$\kappa$ and damping coefficient $\beta$ \cite{metcalf99}. The atoms approach
the axis with the cooling time constant $\tau\simeq\beta/\kappa$. For our
2D~MOT\ parameters $\tau\approx0.5~$\textrm{ms}. Atoms entering the 2D~MOT
with velocity $v<v_{c}$ only contribute to the cold lithium beam if $\tau$ is
less than the residence time $\tau_{\mathrm{res}}$ in the trapping beams
$\left(  \tau<\tau_{\mathrm{res}}\right)  $. In the absence of collisions with
background gas $\tau_{\mathrm{res}}$ is determined by the velocity component
$\left\vert v_{z}\right\vert \lesssim v_{c}a/L$ of the trapable lithium atoms
along the symmetry axis of the 2D~MOT and the entry point in the optical
field. If even the atoms with the shortest residence time can still be cooled,
\textit{i.e.}~for
\begin{equation}
\left\vert v_{c}\right\vert \lesssim\frac{w}{a}\frac{L}{\tau+\tau_{Z}}%
\simeq\frac{L}{\tau}%
\end{equation}
essentially all captured atoms contribute to the cold beam. For
$L=100~\mathrm{mm}$ we calculate with $\tau=1~$\textrm{ms }that this condition
is satisfied for $v_{c}\lesssim100~$\textrm{m/s}, including the experimental
value $v_{c}\approx85~$\textrm{m/s} (see section \ref{section:Discussion}).

\section{Experimental results\label{section:ExperimentalResults}}

\subsection{Oven flux}

To evaluate the merits of the 2D~MOT it is essential to have a reliable
estimate of the input flux from the lithium oven. For this purpose the oven
flux was measured at $T=623\mathrm{~K}$ by observing - in the absence of the
Nd$_{2}$Fe$_{14}$B magnets - the Doppler profile of the hot lithium beam using
a horizontal probe beam with a waist of $1\mathrm{~mm}$ running parallel to
the 2D~MOT axis and back-reflected by a gold-plated mirror (spring-mounted at
the entrance of the DP-channel) as indicated in Fig.\thinspace\ref{fig:vacuum}%
. To avoid optical pumping to dark states the probe intensity was kept at the
low value of $\sim0.018\,I_{sat}$. With a thermal velocity of $\bar{v}%
_{th}=1500~$\textrm{m/s} the interaction time is $1.3$~$\mu$s and the
scattering rate is estimated to be $0.4$ photons per atom. The effect of small
fluctuations in the intensity of the probe laser was suppressed by measuring
the intensity of the probe beam relative to that of a reference beam
originating from the same laser diode. Both the probe beam and the reference
beam were measured with Texas Instruments OPT101 photodiodes.
\begin{figure}
[ptb]
\begin{center}
\includegraphics[
trim=0.199241in 0.225427in 0.249666in 0.455971in,
natheight=3.410400in,
natwidth=4.099600in,
height=6.1527cm,
width=8.2066cm
]%
{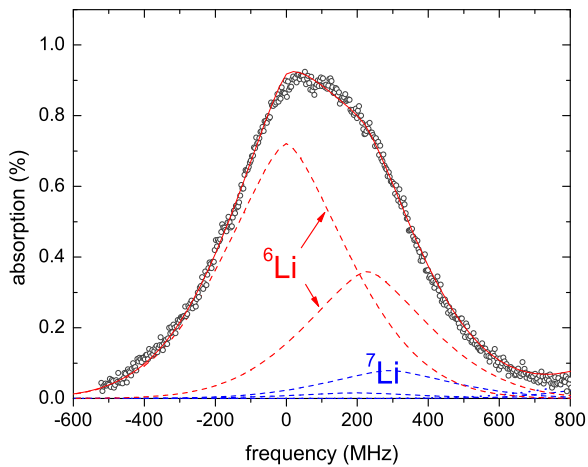}%
\caption{The transverse Doppler profile of the hot lithium flux emerging from
the oven as measured with the horizontal probe beam indicated in
Fig.\thinspace\ref{fig:vacuum}. The calculated profile (solid line) is the sum
of six overlapping Doppler broadened transitions (dotted lines), two of which
have a maximum outside the frequency range shown (see text). Only the
amplitude has been fitted presuming the measured oven temperature $T=623$ K
and $^{6}$Li-abundance of $74\%$. }%
\label{fig:transverseDoppler}%
\end{center}
\end{figure}
The observed Doppler profile is shown in Fig.\thinspace
\ref{fig:transverseDoppler}. The solid line represents a fit of the calculated
Doppler profile for the oven temperature $T=623\mathrm{~K}$ and presuming the
measured $^{6}$Li-abundance. The solid line is the sum of six overlapping
Doppler broadened lines (dotted lines). The unusual lineshapes reflect the
clipping profile of the oven tube. The two large peaks at $0$ and
$228~$\textrm{MHz} correspond to the trapping and repumping transitions in
$^{6}$Li, respectively. Analogously the other four peaks at $199,291,1002$ and
$1094~$\textrm{MHz} are for the $F=2\rightarrow F^{\prime}=1,2$ and
$F=1\rightarrow F^{\prime}=1,2$ transitions of the $D_{1}$ line of $^{7}$Li
\cite{das07}. The best fit is obtained for $\Phi_{s}=8(3)\times10^{13}%
\,$s$^{-1}$, where the error reflects our estimate of systematic
uncertainties. This result overlaps with the value $\Phi_{s}=8_{-3.2}%
^{+6.4}\times10^{13}$\thinspace s$^{-1}$ calculated with Eq.\thinspace
(\ref{eq:Phis}) of the semi-empirical model starting from the oven
temperature.%
\begin{figure}
[ptb]
\begin{center}
\includegraphics[
trim=0.217949in 0.217596in 0.274178in 0.300857in,
natheight=2.660100in,
natwidth=2.488000in,
height=8.1473cm,
width=7.5915cm
]%
{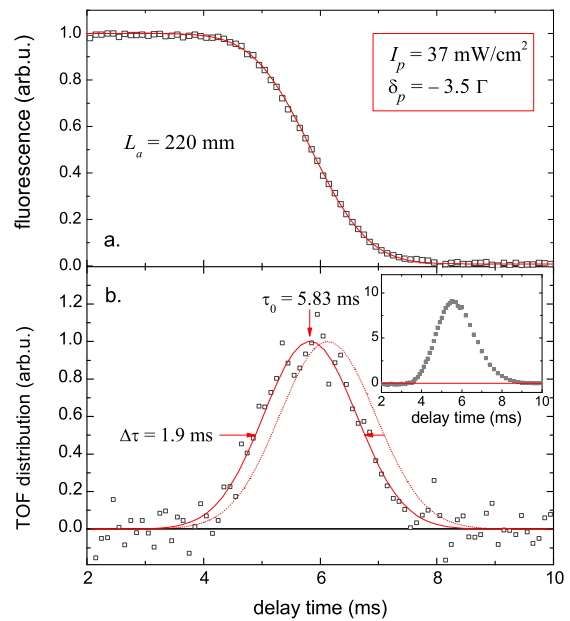}%
\caption{a.) Typical fluorescence decay curve as a function of the probe delay
time. The solid line is a fitted error function. Each datapoint represents the
average over $200$ cycles taken over a period of $6.7$~s. b.) Derivative of
the same data. The dashed line represents the true TOF-distribution
(normalized to the same peak height) as calculated with the model presented in
the text. The inset shows a TOF distribution as measured with a pulsed push
beam. }%
\label{fig:tof}%
\end{center}
\end{figure}

\subsection{Fluorescence detection - TOF
distribution\label{subsection:Fluorescence}}

We probe the intensity of the cold $^{6}$Li beam in the middle of the main
vacuum chamber by measuring the fluorescence after flashing a sheet of
resonant laser light (knife-edge defined: $d=1\mathrm{~mm}$ thick and
$h=5\mathrm{~mm}$ high) propagating horizontally through the middle of the UHV
chamber orthogonal to the beam axis at position $z=L_{a}=220~\mathrm{mm}$
downstream from the entry point of the DP-channel. The fluorescence flash is
imaged vertically as a stripe onto a CCD camera. The length of the stripe
provides information about the divergence of the beam. To remove stray-light
fluctuations the integrated signal from the pixel area containing the stripe
image is divided by the background signal from a reference area. For the probe
beam we use $0.5$~ms flashes of $0.3\mathrm{~W\,cm}^{-2}$ in a ratio of
$1:1.5$ trap/repump light at zero detuning. The beam is retroreflected to
prevent the atoms from being pushed out of resonance.

Velocity characterization of the cold $^{6}$Li beam is done with a
time-of-flight (TOF) method. For this purpose the beam is periodically
interrupted at typically $30$ Hz repetition rate with a resonant
$0.6\mathrm{~W\,cm}^{-2}$ `plug' laser ($2:1$ trap/repump light) deflecting
the atoms near the entrance of the DP-channel. From the decay of the
fluorescence signal $\phi_{\mathrm{fl}}$ as a function time (see
Fig.\thinspace\ref{fig:tof}) we obtain the apparent TOF-distribution, which is
proportional to $d\phi_{\mathrm{fl}}/d\tau$ and can be transformed into the
axial velocity distribution using the flight distance of $220~\mathrm{mm}$. In
a typical measurement we average over $200$ cycles to reach a proper
signal/noise ratio also for small fluxes traversing the light sheet at high velocity.

The procedure is illustrated in Fig.\thinspace\ref{fig:tof} for a push-beam
intensity of $I_{p}=37\mathrm{~mW\,cm}^{-2}$ and a detuning $\delta
_{p}=-3.5\,\Gamma$. Note that the derivative of $\phi_{\mathrm{fl}}$ can be
nicely described by the gaussian function
\begin{equation}
d\phi_{\mathrm{fl}}/d\tau=(\pi^{1/2}\Delta\tau)^{-1}\exp[-(\tau-\tau_{0}%
)^{2}/\Delta\tau^{2}],
\end{equation}
where $\tau_{0}=5.83~\mathrm{ms}$ is the mean apparent arrival time and
$1.67\,\Delta\tau=1.9~\mathrm{ms}$ is the full width at half maximum (FWHM).
The absence of arrival times shorter than $3~\mathrm{ms}$ reflects the absence
of atoms with velocities $v_{z}\gtrsim70~$\textrm{m/s}. This absence of `hot'
flux was verified up to $4$~$\mathrm{km/s}$ and was anticipated because the
cold beam is pushed horizontally out of the 2D~MOT, \textit{i.e.}~orthogonally
to the hot flux from the oven. The observed relative spread $\Delta\tau
/\tau_{0}\approx0.2$ is insensitive to the push-beam intensity and comparable
to the instrumental resolution for the shortest flight times investigated
($\tau_{0}=3~\mathrm{ms})$. The value of $\tau_{0}$ is entirely determined by
the properties of the push beam and insensitive to other 2D~MOT parameters.
This behavior was previously also observed in other 2D~MOT systems
\cite{inguscio06,pfau02}. Since optical pumping to different hyperfine states
takes only a few optical cycles in $^{6}$Li and $L_{a}/\tau_{0}=38~$%
\textrm{m/s} corresponds to $\sim380$ photon recoils, the atoms must have been
accelerated to their final velocity still within reach of the repump light,
\textit{i.e.~}inside 2D~MOT (the push beam does not contain repumper light).
This limits the acceleration to a well-defined duration of time, which is
consistent with the observed relatively narrow velocity distribution. The
absence of slow atoms is not caused by gravity because for the lowest
velocities measured $\left(  L_{a}/\tau_{0}=22~\mathrm{m/s}\right)  $ the
gravitational drop is only $0.5\mathrm{~mm}$, less than half the height
$\left(  h/2=2.5\text{\textrm{~mm}}\right)  $ of the light sheet.

To relate the fluorescence signal $\phi_{\mathrm{fl}}$ to the velocity
distribution in the atomic beam we have to account for the detection
efficiency, which is inversely proportional to the velocity of the atoms and
depends on the divergence of the beam. For this purpose we approximate the
beam spot at the position of the light sheet $\left(  z=L_{a}\right)  $ by a
gaussian profile with $1/e$-radius $R$. The fraction $\chi_{\mathrm{fl}}$ of
the beam giving rise to fluorescence is obtained by integrating the normalized
gaussian beam profile in horizontal and vertical direction over the surface
area of the light sheet,%
\begin{equation}
\chi_{\mathrm{fl}}=\operatorname{erf}(h/2R)\operatorname{erf}(S_{0}/R),
\label{eq:integral-1}%
\end{equation}
where $S_{0}=4.5\mathrm{~mm}$ is the radius of the optical field of view. Here
we neglected some clipping by the DP-channel. Note that the divergence angle
$\zeta$ of the cold beam equals the ratio of transverse to axial velocity of
the atoms, $\zeta=R/L_{a}=v_{t}/v_{z}$. The length of the fluorescence stripe
was found to vary only slightly with the intensity of the push beam. This sets
a lower bound on the beam divergence, $S_{0}/R\lesssim1$ for $v_{z}%
=70~$\textrm{m/s} and on the characteristic transverse velocity, $v_{t}%
\gtrsim1.4~$\textrm{m/s}. Since $h/2R\ll S_{0}/R\lesssim1$ for all velocities
studied Eq.\thinspace(\ref{eq:integral-1}) can be written in the form%
\begin{equation}
\chi_{\mathrm{fl}}(v_{z}/v_{t})\simeq\gamma v_{z}/v_{t}\operatorname{erf}%
\left(  \eta v_{z}/v_{t}\right)  , \label{eq:integral-2}%
\end{equation}
where $\eta=S_{0}/L_{a}=0.02$ is the view angle and $\gamma=h/2L_{a}=0.011$
the vertical acceptance angle.
\begin{figure}
[ptb]
\begin{center}
\includegraphics[
trim=0.223801in 0.205758in 0.209195in 0.393544in,
natheight=3.667700in,
natwidth=4.711600in,
height=5.7925cm,
width=8.055cm
]%
{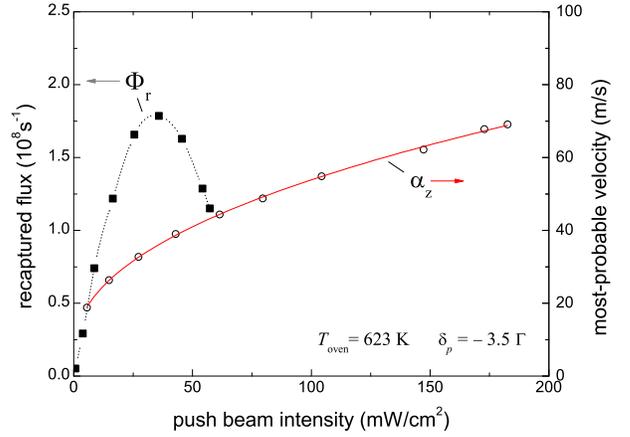}%
\caption{Recapture rate into the 3D~MOT (solid squares - left scale) and the
most probable axial velocity $\left(  \alpha_{z}\right)  $ of the cold atomic
beam (open circles - right scale) both as a function of the push beam
intensity. The drawn lines provide a guide to the eye.}%
\label{fig:velVsPush}%
\end{center}
\end{figure}

The fluorescence decay signal $\phi_{\mathrm{fl}}$ can be expressed in the
form%
\begin{equation}
\phi_{\mathrm{fl}}(\tau)\sim\int_{0}^{L_{a}/\tau}\frac{\chi_{\mathrm{fl}%
}(v_{z}/v_{t})}{v_{z}}\phi_{0}\left(  v_{z},\alpha_{z}\right)  \,dv_{z},
\end{equation}
where $\phi_{0}\left(  v_{z},\alpha_{z}\right)  $ is the normalized axial
velocity distribution with $\alpha_{z}$ representing the most-probable
velocity in the beam, and $L_{a}/\tau$ the velocity of the fastest atoms still
arriving at the detector after delay time $\tau$. Hence, the transformation
between the beam property $\phi_{0}\left(  v_{z},\alpha_{z}\right)  $ and the
observed fluorescence decay is given by%
\begin{equation}
\phi_{\mathrm{TOF}}\left(  \tau\right)  =\phi_{0}\left(  L_{a}/\tau\right)
\varpropto-\left(  \tau/\chi_{\mathrm{fl}}\right)  d\phi_{\mathrm{fl}}/d\tau.
\label{eq:flux}%
\end{equation}
Here $\phi_{\mathrm{TOF}}\left(  \tau\right)  $ represents the distribution of
flight times in the beam. For $\Delta\tau/\tau\ll1$ the prefactor $\left(
\tau/\chi_{\mathrm{fl}}\right)  $ causes the distribution $d\phi_{\mathrm{fl}%
}/d\tau$ to shift to larger delay times but its shape remains well-described
by a gaussian. In our case the shift is $5\%$ $\left(  \tau_{\max}%
\simeq1.05\,\tau_{0}\right)  $ as indicated by the dotted line in
Fig.\thinspace\ref{fig:tof}. Hence, the most-probable velocity in the beam is
given by $\alpha_{z}\simeq0.95\,L_{a}/\tau_{0}$. For the example of
Fig.\thinspace\ref{fig:tof} we calculate $\alpha_{z}=36~\mathrm{m/s}$ with a
FWHM of $11~\mathrm{m/s}$. We have observed a ten-fold increase in
$\phi_{\mathrm{TOF}}\left(  \tau\right)  $ at constant average flux by pulsing
the push beam (see inset in Fig.\thinspace\ref{fig:tof}). This indicates that
the 2D~MOT is not limited by its density when the push beam is continuously
on. The most-probable velocity $\alpha_{z}$ was found to be the same for
pulsed and continuous operation. The experimental results for $\alpha_{z}$ as
a function of the push beam intensity are shown as the open circles in
Fig.\thinspace\ref{fig:velVsPush}. Varying the push-beam intensity $I_{p}$
over the range $5-180~\mathrm{mW\,cm}^{-2}$ we found $\alpha_{z}$ to increase
from $18-70~\mathrm{m/s}$.

\subsection{Beam flux - dependence on push beam}

The flux of the cold atomic beam is investigated as a function of the
push-beam intensity $\left(  I_{p}\right)  $ by recapture into the 3D~MOT. The
results are shown as the solid squares in Fig.\thinspace\ref{fig:velVsPush}.
First of all we note that in the absence of the push beam the flux arriving at
the recapture MOT is very small. Under these conditions the 2D~MOT performance
is very sensitive to the alignment of the quadrupole field, the MOT beams and
the repumper. This low flux is attributed to the horizontal orientation of the
beam axis, orthogonal to the direction of the hot flux from the oven. In view
of this symmetry the trapped atoms have an axial velocity distribution
centered around zero. Only the atoms with axial velocity $v_{z}\gtrsim
5~$\textrm{m/s} will reach the capture volume of the 3D~MOT. Slower atoms drop
below the trapping region as a result of gravity. High-field-seeking atoms
will be deflected away from the recapture MOT by the quadrupole field outside
the 2D~MOT for axial velocities $v_{z}\lesssim10~$\textrm{m/s}. Atoms with
axial velocity $v_{z}\gtrsim0.1\,v_{c}\approx8.5~$\textrm{m/s} are absent due
to clipping by the oven tube ($v_{c}$ is the capture velocity of the 2D~MOT).

As an aside we point out that by inclining the axis of the oven tube toward
the beam axis direction it should be possible to realize a high flux cold beam
with an axial velocity proportional to the inclination angle and without any
(near)resonant light co-propagating with the atomic beam into the UHV chamber.
In a more practical solution this may be realized by not retroreflecting the
2D-MOT beams but tilting them so that the average $\mathbf{k}$-vector points
along the cold beam axis.%

\begin{figure}
[ptb]
\begin{center}
\includegraphics[
natheight=8.028100in,
natwidth=9.374500in,
height=6.6843cm,
width=7.7914cm
]%
{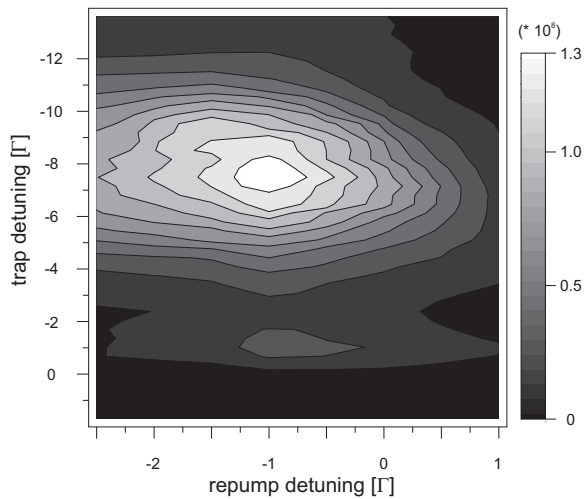}%
\caption{The 3D~MOT loading rate as a function of the 2D~MOT trap and repump
detunings for an oven teperature $T=623(12)~\mathrm{K}$. These measurements
were performed with maximum power for the 2D~MOT trap and repump beams as
given in Table \ref{table-1}. We find the maximum flux of $1.3\times10^{8}%
\,$\textrm{s}$^{-1}$at a trap detuning of $\delta_{t}=-7.5\,\Gamma$ and
$\delta_{r}=-1\,\Gamma$.}%
\label{fig:2DtrapRep}%
\end{center}
\end{figure}

Measuring the loading rate $\Phi_{r}$ in the 3D~MOT we obtain the `useful'
flux of the cold $^{6}$Li beam. The rate is obtained from the leading slope of
the loading curve, observing the 3D~MOT fluorescence as a function of time
using a CCD camera. This fluorescence is calibrated against an absorption
image taken immediately after switching-off the 3D~MOT. The measured rate
$\Phi_{r}$ represents a lower limit for the flux emerging from the 2D~MOT.
Fig.\thinspace\ref{fig:velVsPush} shows that $\Phi_{r}$ increases steeply
until it reaches a maximum at $I_{p}\approx34~$mW\thinspace cm$^{-2}$. Further
increase of the push-beam intensity causes the loading rate to decrease. This
is attributed to the finite capture velocity of the 3D~MOT (see section
\ref{section:Discussion}). For the data shown in Fig.\thinspace
\ref{fig:velVsPush} we used for the 3D~MOT a magnetic field gradient of
$0.19\mathrm{~T/m}$, $10\mathrm{~mW}$ trapping light per beam at a detuning of
$-6$ $\Gamma$ and $11\mathrm{~mW}$ repumping light per beam at a detuning of
$-3.5$ $\Gamma$. Both colors are distributed over six beams clipped at their
beam waist of $9\mathrm{~mm}$, thus defining the acceptance radius
$R_{a}=9\mathrm{~mm}$ of the 3D~MOT.\begin{table}[b]
\caption{Experimental parameters for optimal performance of the Li 2D~MOT.}%
\label{table-1}%
\begin{tabular}
[c]{cccc}\hline
parameter & trap & repump & push\\\hline\hline
detuning $\delta$ & -7.5 $\Gamma$ & -1 $\Gamma$ & -3.5 $\Gamma$\\
power per beam & 50 mW & 48 mW & 0.8 mW\\
waist ($1/e^{2}$ radius) & 9 & 9 & 1.2\\\hline
gradient & 50 G/cm &  & \\
oven temperature & 623 K &  & \\
most-probable velocity & 36 m/s &  & \\
FWHM\ of velocity distribution & 11 m/s &  & \\\hline
\end{tabular}
\end{table}%
\begin{figure}
[ptb]
\begin{center}
\includegraphics[
trim=0.204222in 0.200000in 0.350941in 0.438384in,
natheight=3.367000in,
natwidth=4.228200in,
height=5.6783cm,
width=7.6267cm
]%
{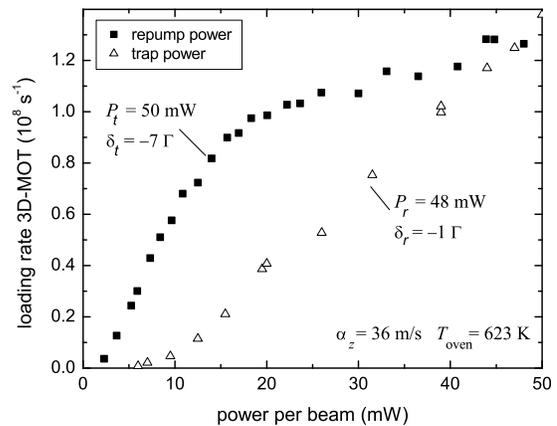}%
\caption{The 3D~MOT loading rate as a function of the trap and repump laser
powers (power per beam). Note that the 2D~MOT is operated in a retroreflected
configuration.}%
\label{fig:2Dpowers}%
\end{center}
\end{figure}

\subsection{Beam flux - dependence on 2D~MOT}

We have optimized the total flux by varying both the trap and the repump
detuning. For these measurements the laser power of the trap and repump lasers
were set to their maximum values of $100\mathrm{~mW}$ and $94\mathrm{~mW}$,
respectively. The results are shown as a contour diagram in Fig.\thinspace
\ref{fig:2DtrapRep}. The maximum flux is observed when the trap laser is far
detuned ($\delta_{t}=-7.5\Gamma$) and the repump laser is close to resonance
($\delta_{r}=-1\Gamma$). We observe a small local maximum in flux if the trap
laser is tuned close to resonance ($\delta_{t}=-1\Gamma$). We attribute this
to better beam collimation because the 2D~MOT is expected to be transversely
colder when operated close to resonance \cite{shimizu93,grimm98}. Apparently
the advantage of better collimation cannot compensate loss in 2D~MOT capture efficiency.

With optimized detunings we measured $\Phi_{r}$ as a function of the available
optical power in the 2D~MOT trap $(P_{t})$ and repump $\left(  P_{r}\right)  $
beams. For this purpose either the trapping power is kept constant at
$P_{t}\approx50\mathrm{~mW}$ per beam and $P_{r}$ is varied or the repumping
power is kept constant at $P_{r}\approx48\mathrm{~mW}$ per beam and $P_{t}$ is
varied. As is shown in Fig.~\ref{fig:2Dpowers} the loading rate increases
linearly with $P_{t}$ for $P_{t}\gtrsim8\mathrm{~mW}$, whereas $\Phi_{r}$
increases linearly with $P_{r}$ for $P_{r}\gtrsim2\mathrm{~mW}$ until it
levels off for $P_{r}\gtrsim18\mathrm{~mW}$. The experimental parameters for
optimal source performance are collected in Table \ref{table-1}. The output
flux was reproducible to within $30\%$ depending on the 2D~MOT alignment.

\subsection{Beam flux - dependence on oven temperature}

\begin{figure}
[ptb]
\begin{center}
\includegraphics[
trim=0.180386in 0.206060in 0.180819in 0.363636in,
natheight=3.367000in,
natwidth=4.325800in,
height=5.6458cm,
width=7.9731cm
]%
{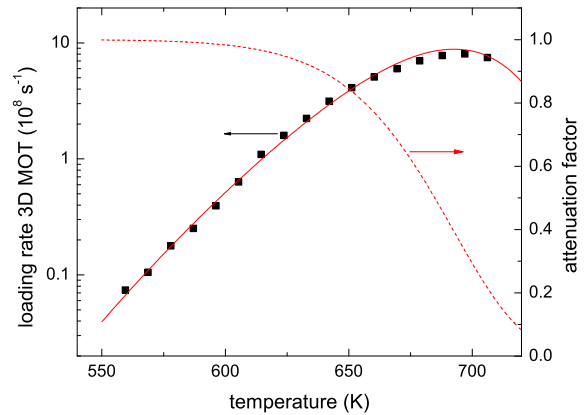}%
\caption{The 3D~MOT loading rate as a function of oven temperature (solid
squares - left scale). The solid line shows a fit of the model presented in
section \ref{section:Discussion}. The loading rate reaches a maximum of
$8\times10^{8}\,\mathrm{s^{-1}}$ at $T\approx700~\mathrm{K}$ as a result of
beam attenuation by hot background vapor. The calculated attenuation factor is
shown as the dashed line (right scale).}%
\label{fig:loadingvstemp}%
\end{center}
\end{figure}

Fig.\thinspace\ref{fig:loadingvstemp} shows the loading rate as a function of
the oven temperature. At low temperatures the loading rate increases
exponentially with the oven temperature. This reflects the exponential
increase of the effusive flux from the oven. Above $T\approx650$~\textrm{K} a
loss mechanism sets in. This limits further increase of the flux until at
$T\approx700~$\textrm{K} the cold atomic flux reaches its maximum value,
corresponding to a loading rate of $\Phi_{r}=8(3)\times10^{8}$\thinspace
s$^{-1}$ into the 3D~MOT. The error reflects our best estimate of systematic
uncertainties. As will be will be discussed in section
\ref{section:Discussion} the losses are attributed to knock-out collisions in
the effusive beam emerging from the oven. The dotted line shows the fraction
of atoms surviving the loss mechanism.%

\section{Discussion\label{section:Discussion}}

\subsection{Recapture in the 3D~MOT\label{subsection:recapture}}

\begin{figure}
[ptb]
\begin{center}
\includegraphics[
trim=0.196092in 0.215352in 0.196511in 0.374567in,
natheight=3.282800in,
natwidth=4.190000in,
height=5.1818cm,
width=7.284cm
]%
{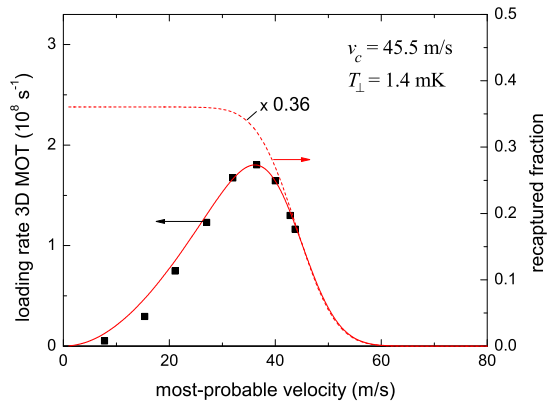}%
\caption{Loading rate 3D-MOT as a function of the most-probable velocity
$\alpha_{z}$ in the beam (black squares - left scale). The drawn line
represents the best fit to the data of the recapture model described in the
text for $v_{c}=45.5\mathrm{~m/s}$ (right scale). The result for zero beam
divergence is shown as the dotted line, scaled down with a factor $0.36$ for
convenience of comparison. }%
\label{figRate-vs-PeakVelocity}%
\end{center}
\end{figure}

To analyze the performance of the 2D~MOT source we define the overall
efficiency parameter $\chi$ as the ratio of the 3D~MOT loading rate $\Phi_{r}$
and the maximum capturable flux $\Phi_{c}$ from the oven,%
\[
\Phi_{r}=\chi\Phi_{c}.
\]
This efficiency is determined by the capture efficiencies of the 2D and 3D~MOT
as well as the transfer efficiency $\chi_{t}$ related to the divergence of the
atomic beam. To determine $\chi_{t}$ as well as the capture velocity $v_{c}$
we replotted the data of Fig.\thinspace\ref{fig:velVsPush} in the form of
Fig.\thinspace\ref{figRate-vs-PeakVelocity}, showing the capture rate
$\Phi_{r}$ in the 3D~MOT as a function of the most-probable axial velocity
$\alpha_{z}$ in the cold atomic beam. Like in subsection
\ref{subsection:Fluorescence} we approximate the atomic beam profile at the
position of the 3D~MOT $\left(  z=L_{a}=220\mathrm{~mm}\right)  $ by the
gaussian profile with $1/e$-radius $R$. The transfer efficiency is obtained by
integrating the normalized profile from $r=0$ on the beam axis to the
acceptance radius $r=R_{a}=9\mathrm{~mm}$ of the 3D~MOT,%
\begin{equation}
\chi_{t}(x_{a})\simeq2\int_{0}^{x_{a}}\left(  1-x/x_{0}\right)  e^{-x^{2}}xdx.
\label{eq:TransferEfficiency}%
\end{equation}
Here $x=r/R$, $x_{a}=R_{a}/R$ and $x_{0}=R_{0}/R$. The factor $\left(
1-x/x_{0}\right)  $ represents the conical approximation to the trapezoidal
clipping profile imposed by the DP-channel, where $R_{0}=19\mathrm{~mm}$ marks
the edge of the dark shadow. Defining the 3D~MOT acceptance angle
$\alpha=R_{a}/L_{a}$ and velocity ratio $\tilde{v}_{z}\equiv v_{z}%
/v_{t}=1/\zeta$ we write compactly $x_{a}=\alpha\tilde{v}_{z}$. Similarly we
define the clipping angle $\beta=R_{0}/L_{a}$ and write $x_{0}=\beta\tilde
{v}_{z}$. Substituting the expressions for $x_{a}$ and $x_{0}$ into
Eq.\thinspace(\ref{eq:TransferEfficiency}) and evaluating the integral we
obtain for the transfer efficiency%
\begin{equation}
\chi_{t}(\tilde{v}_{z})=1-(1-\alpha/\beta)e^{-\left(  \alpha\tilde{v}%
_{z}\right)  ^{2}}-\frac{1}{2\beta\tilde{v}_{z}}\sqrt{\pi}\operatorname{erf}%
\left(  \alpha\tilde{v}_{z}\right)  .
\end{equation}
The velocity-averaged transfer efficiency (recaptured fraction) into the
3D~MOT is given by%
\begin{equation}
\bar{\chi}_{t}(\alpha_{z},v_{c})=\int_{0}^{v_{c}}\chi_{t}(v_{z}/v_{t})\phi
_{0}\left(  v_{z},\alpha_{z}\right)  dv_{z}, \label{CapturedFraction}%
\end{equation}
where $\phi_{0}\left(  v_{z},\alpha_{z}\right)  $ is the normalized axial
velocity distribution defined by Eq.\thinspace(\ref{eq:flux}). The solid line
in Fig.\thinspace\ref{figRate-vs-PeakVelocity} is a plot of $\bar{\chi}%
_{t}(\alpha_{z},v_{c})$ for fixed value of $v_{c}$. The position of the
maximum is insensitive for the beam divergence and the best fit is obtained
for a capture velocity of $v_{c}=45.5\mathrm{~\mathrm{m/s}}$. In contrast the
peak height $\bar{\chi}_{\max}$ depends strongly on the beam divergence. Using
the lower limit for the characteristic transverse velocity $\left(
v_{t}\gtrsim1.4\mathrm{~\mathrm{m/s}}\right)  $ we calculate an upper limit
for the recaptured fraction $\bar{\chi}_{\max}\lesssim0.4$. For comparison
also the result for zero beam divergence is shown in the plot (dotted line).%

For the conditions used in experiment, $\partial B/\partial
r=0.19~\mathrm{T/m}$ and $\delta_{L}=-6\Gamma$, we calculate with
Eq.\thinspace(\ref{delta}) $v_{c}=32+5.6\left\vert \delta_{L}/\Gamma
\right\vert \approx66~\mathrm{m/s}$ for $r_{c}=\sqrt{2}R_{a}=12.7\mathrm{~mm}%
$. Apparently the simple 1D model overestimates the capture velocity by some
$50\%$. Because both the 2D~MOT and the 3D~MOT are configured in the
$135^{\circ}$ configuration with respect to the input beam and also
$r_{c}=12.7\mathrm{~mm}$ in both cases we presume a similar overestimate for
the capture velocity of the 2D~MOT. In the latter case we have $\partial
B/\partial r=0.50~\mathrm{T/m}$ and $\delta_{L}=-7.5\Gamma$ and calculate with
Eq.\thinspace(\ref{delta}) $v_{c}=85+5.6\left\vert \delta_{L}/\Gamma
\right\vert \approx127~\mathrm{m/s}$. Presuming somewhat arbitrarily that also
this value overestimates the actual capture velocity by $50\%$ we obtain
$v_{c}\approx85~\mathrm{m/s}$ as a reasonable estimate.

Starting from $\Phi_{s}=8(3)\times10^{13}$\textrm{~s}$^{-1}$ we obtain with
Eq.\thinspace(\ref{eq:PhiMax}) for the theoretical maximum flux $\Phi
_{c}=5(2)\times10^{8}$\textrm{~s}$^{-1}$. With the measured value $\Phi
_{r}=1.8(6)\times10^{8}$\textrm{~s}$^{-1}$ the overall efficiency $\chi
=\Phi_{r}/\Phi_{c}$ is estimated to be $0.2\lesssim\chi\lesssim0.5$. This set
a lower limit on the recaptured fraction, $0.2\lesssim\bar{\chi}_{\max
}\lesssim0.4$, and (using our model) also an upper limit for the
characteristic transverse velocity, $v_{t}\lesssim2.5~$\textrm{m/s.} As the
upper and lower limits more or less coincide our best estimate is
$v_{t}\approx2~$\textrm{m/s}, which corresponds to a transverse 2D~MOT
temperature of $T_{\perp}=1.4~\mathrm{mK}$. The corresponding beam divergence
at optimal recapture for oven temperature $T=623(12)~\mathrm{K}$ is
$\zeta\approx0.05$. For these conditions the brightness of the beam emerging
from the 2D~MOT is calculated to be $\sim2\times10^{11}$\textrm{~sr}$^{-1}%
$\textrm{s}$^{-1}$.

\subsection{Loss mechanisms\label{subsection:LossMechanisms}}

Because $\Phi_{c}\lll\Phi_{s}$ the output from the oven is well characterized
by a small cold flux of capturable atoms overtaken by the hot flux of the full
emittance. Once the hot flux exceeds a critical value we expect the cold flux
to be attenuated by `knock-out' collisions. This depletion of the low velocity
class of atoms is a well-known phenomenon in close-to-effusive beam sources
\cite{estermann47}. Comparing the total flux per unit area just above the
emitting surface, $\Phi_{tot}/A\approx6.5\times10^{15}$\textrm{\thinspace
s}$^{-1}$\textrm{cm}$^{-2}$, with the flux per unit area in the capture region
$\Phi_{s}/A\approx4\times10^{13}$\textrm{\thinspace s}$^{-1}$\textrm{cm}%
$^{-2}$, we expect these knock-out collisions to occur primarily in the first
few centimeters of the expanding beam. Once the atoms enter the 2D MOT the
cross section increases because optically excited atoms interact resonantly
with the hot background flux \cite{margenau71}.

To model the attenuation we calculate the collision rate of an atom moving at
velocity $v_{c}$ along the symmetry axis at position $l$ above the oven exit
orifice with atoms from the hot background flux moving at typically the
average velocity $\bar{v}\gg v_{c}$,%
\begin{equation}
\dot{\Phi}/\Phi=\tfrac{1}{2}\sigma_{6}n_{s}\int_{0}^{\theta_{0}(l)}v_{r}%
\sin\theta d\theta.
\end{equation}
Here $\theta$ is the emission angle of the fast moving atoms with respect to
the symmetry axis, $\tan\theta_{0}=a/l$, $\sigma_{6}$ is the knock-out cross
section and $v_{r}=(\bar{v}^{2}+v_{c}^{2}-2v_{c}\bar{v}\cos\theta)^{1/2}%
\simeq\bar{v}$ is the relative velocity of the colliding atoms
\cite{estermann47}. Using the substitution $dl=v_{c}dt$ we can solve the
differential equation under the boundary condition $\Phi(l)=\Phi_{c}$ at $l=0$
and obtain
\begin{equation}
\Phi(L)\simeq\Phi_{c}\exp[-\tfrac{1}{2}\sigma_{6}n_{s}\left(  \bar{v}%
/v_{c}\right)  \int_{0}^{L}(1-\cos\theta_{0})dl], \label{eq:AttenuatedFlux}%
\end{equation}
where $\cos\theta_{0}=l/(l^{2}+a^{2})^{1/2}$. In this model the density in the
oven is taken to be uniform. Because for $l\gg a$ the collision probability
vanishes we may freely extend the integral to infinity, $\int_{0}^{\infty
}(1-\cos\theta_{0})dl=a$. Hence, at the entrance of the 2D MOT the attenuated
flux is given by%
\begin{equation}
\Phi_{in}=\lim_{L\rightarrow\infty}\Phi(L)\simeq\Phi_{c}\exp(-\sigma_{6}%
n_{s}\bar{v}\tau_{6}), \label{eq:LossesC6}%
\end{equation}
where $\tau_{6}=a/2v_{c}\approx47\mathrm{~\mu s}$ is the characteristic
duration of the attenuation process.

To estimate $\sigma_{6}$ we take the approach of ref.~\cite{steane92} and
consider a slow atom moving at the capture velocity $v_{c}$ along the
symmetry-axis from the oven towards the capture region. Fast atoms flying-by
with the thermal velocity $\bar{v}$ will give rise to momentum transfer as a
result of Van der Waals interaction. As this happens most frequently close to
the oven even a small momentum transfer $\Delta p\lesssim0.1\,mv_{c}$ suffices
to kick the atoms out of the capture cone $\Omega_{c}$. Because $\bar{v}\gg
v_{c}$ the trajectory of the fast atom is hardly affected and the momentum
transfer to the cold atom can be calculated by integrating the transverse
component of the Van der Waals force over time, $\Delta p=\frac{1}{2}%
\int_{-\infty}^{\infty}F_{\bot}(t)dt$. Here $F(r)=6C_{6}/r^{7}$ with $r$ the
radial distance between the colliding atoms and $C_{6}=1389\,a_{0}^{6}\,E_{h}$
the Van der Waals coefficient \cite{dalgarno01} with $a_{0}\approx
0.529\times10^{-10}\,\mathrm{m}$ the Bohr radius and $E_{h}\approx
4.36\times10^{-18}\,\mathrm{J}$ the Hartree energy. Changing from the time
variable $t$ to the angular variable $\theta$ using $\tan\theta=\bar{v}t/b$,
where $b$ is the distance of closest approach, we obtain using $F_{\bot}%
=F\cos\theta$ and $\cos\theta=b/r$,
\begin{equation}
\Delta p=\frac{6C_{c}}{2\bar{v}b^{6}}\int_{-\pi/2}^{\pi/2}\cos^{6}\theta
d\theta=\frac{C_{6}}{\bar{v}b^{6}}\frac{15\pi}{16}.
\end{equation}
The critical distance of closest approach for which the atoms are just
scattered outside the capture cone $\Omega_{c}$ is given by
\begin{equation}
b_{6}\simeq1.8\left(  C_{6}/mv_{c}\bar{v}\right)  ^{1/6}. \label{eq:b}%
\end{equation}
Note that this quantity depends only very weakly on the precise values of
$v_{c}\ $and $\bar{v}$. For $v_{c}\approx85~\mathrm{m/s}$ and temperatures in
the range $600\lesssim T\lesssim700\mathrm{~K}$ we calculate for the knock-out
cross section $\sigma_{6}=\pi b_{6}^{2}\approx4.4\times10^{-14}~\mathrm{cm^{2}%
}$. Note that, in contrast to `knock-out' collisions, `knock-in' collisions
are rare. The steep dependence of $\Delta p$ on $b$ implies that most of the
atoms scattered outside the acceptance cone scatter over much larger angles
than the minimum angle required for knock-out. Thus scattered atoms typically
hit the wall of the oven tube and stick, rather than giving rise to knock-in.

Along the same lines we estimate the momentum transfer by resonant collisions
inside the 2D~MOT. As the relative velocities are large and the typical
collision time is much shorter than the lifetime of the atoms in the excited
state we may use again the classical scattering model discussed above. In the
present case the critical distance of closest approach corresponds to momentum
transfer just exceeding the escape value from the 2D MOT, $mv\gtrsim mv_{c}$
\cite{steane92}. Neglecting the direction of the transition dipole the
resonant-dipole force can be approximated by $F(r)=3C_{3}/r^{4}$, where the
$C_{3}$ coefficient is defined as \cite{margenau71,fontana61}%
\begin{equation}
C_{3}=e^{2}a_{0}^{2}D_{eg}^{2}/4\pi\varepsilon_{0}=3.7\times10^{-48}%
\,\mathrm{J\,m}^{3}.
\end{equation}
Here $e\approx1.60\times10^{-19}\,\mathrm{C}$ is the elementary charge,
$\varepsilon_{0}\approx8.85\times10^{-12}\,\mathrm{Fm}^{\mathrm{-1}}$ the
electric constant and $D_{eg}=2.4~\mathrm{a.u.}$the transition dipole moment
for the $2s\rightarrow2p$ transition in Li \cite{metcalf99}. The corresponding
critical distance of closest approach is in this case
\begin{equation}
b_{3}\simeq1.6\left(  C_{3}/mv_{c}\bar{v}\right)  ^{1/3}. \label{eq.b3}%
\end{equation}
For $v_{c}\approx85~\mathrm{m/s}$ and temperatures in the range $600\lesssim
T\lesssim700\mathrm{~K}$ we calculate for the resonant cross section
$\sigma_{3}=\pi b_{3}^{2}\approx1.6\times10^{-13}~\mathrm{cm^{2}}$. Accounting
for the knock-out probability of trapped atoms the loading rate into the 3D
MOT can be written as
\begin{equation}
\Phi_{r}=\bar{\chi}_{t}\Phi_{in}\exp[-\sigma_{3}\tau_{\mathrm{res}}\Phi
_{s}/A_{c}]. \label{eq:LossesC3}%
\end{equation}

Combining Eqs.\thinspace(\ref{eq:Phis}) and (\ref{eq:PhiMax}) with the $C_{6}$
and $C_{3}$ loss exponents of Eqs.\thinspace(\ref{eq:LossesC6}) and
(\ref{eq:LossesC3}) and introducing the characteristic attenuation time
$\tau_{3}=A\tau_{\mathrm{res}}/4\pi L^{2}$ we obtain the following expression
for the 3D~MOT loading rate%
\begin{equation}
\Phi_{r}\simeq\bar{\chi}_{t}a_{6}n_{s}\bar{v}\,A\left(  \frac{v_{c}}{\alpha
}\right)  ^{4}\frac{\Omega_{c}}{8\pi}\exp[-n_{s}\bar{v}\left(  \sigma_{6}%
\tau_{6}+\sigma_{3}\tau_{3}\right)  ], \label{eq:coldFluxFnT}%
\end{equation}
Using $\tau_{\mathrm{res}}=1~\mathrm{ms}$ we have $\tau_{3}\approx
1.6\mathrm{~\mu s}$. Note that only $\alpha$, $\bar{v}$ and $n_{s}$ are
sensitive for the oven temperature. A best fit of Eq.\thinspace
(\ref{eq:coldFluxFnT}) to the data using $\bar{\chi}_{t}$ and $n_{s}$ (at
$T=623~\mathrm{K}$) as free parameters is shown as the solid line in
Fig.\thinspace\ref{fig:loadingvstemp}. The fit shown is obtained for
$\bar{\chi}_{t}=0.33$ and $n_{s}=1.5\times10^{17}\,\mathrm{m}^{-3}$ at
$T=623~\mathrm{K}$, which are both within the error limits given for these
quantities. Thus also the position of the maximum confirms our model. As the
result obtained for $\bar{\chi}_{t}$ strongly anti-correlates with the value
presumed for $v_{c}$ we cannot improve upon the estimate $\chi_{\max}%
=30\pm10\%$ already given in subsection \ref{subsection:recapture}.

Interestingly, comparing the two loss mechanisms we find $\sigma_{3}\tau
_{3}/\sigma_{6}\tau_{6}\approx0.1$, which shows that the resonance mechanism,
dominating the background losses in the VCMOT \cite{steane92,dieckmann98}, is
of minor importance in the present case. Since the output flux scales like
$(v_{c}/\alpha)^{4}$ an obvious way to increase the output of MOT sources is
to increase the capture velocity. Doubling the waist of the 2D~MOT beams in
the $xy$ plane (see Fig.\thinspace1) in order to increase the capture radius
we find with Eq.\thinspace(\ref{eq:vmax2}) that the capture velocity increases
by $\sqrt{2}$ and the output by a factor $4$.\ In addition, since $\tau_{6}$
scales like $1/v_{c}$ the beam attenuation decreases slightly.

\subsection{Comparison with Zeeman slowers \label{subsection:comparison}}

In several respects the 2D~MOT source demonstrated in this paper represents an
interesting alternative for the Zeeman slower. First of all the source yields
a large controllable output flux of up to $3\times10^{9}~\mathrm{s^{-1}}$,
comparable to fluxes typically achieved in lithium Zeeman slowers. The
transverse temperature of the source is low $\left(  1.4~\mathrm{mK}\right)  $
which makes it possible to recapture as much as $30\%$ in a 3D~MOT
$220~\mathrm{mm}$ downstream from the source. In contrast to Zeeman slowers,
the 2D~MOT source yields a clean and monochromatic cold atomic beam of which
the most probable velocity can be varied over a wide range of velocities with
the aid of a push beam. Permanent magnets for the creation of the quadrupole
field add to the simplicity of the design. The resulting source is more
compact than a typical Zeeman slower and is still capable of loading $10^{10}$
atoms in a 3D~MOT.

Importantly, the 2D~MOT principle works equally well with light atoms as with
more heavy atoms like K, Rb and Cs. This shows that, like the Zeeman slower,
also the 2D~MOT beam source has a wide applicability. In cases with a sizable
vapor pressure at room temperature the source will act as a VCMOT. As an
example of a system for which a 2D~MOT has not yet been realized we briefly
discuss the case of Na. In this case the gradient of the quadrupole field
should be scaled down proportional to $m^{1/2}$ in accordance with
Eq.\thinspace(\ref{eq:-m1/2}) to obtain the optimum value $\partial B/\partial
r\approx0.25~\mathrm{T/m}$. In view of Eq.\thinspace(\ref{eq:vmax2}) the
capture velocity scales down with the same factor. Using Eq.\thinspace
(\ref{eq:coldFluxFnT}) we calculate for Na a maximum total output flux of
$4\times10^{9}~\mathrm{s^{-1}}$ for an oven temperature $T\approx
471~\mathrm{K}$. This output is lower than realized with Zeeman slowers but
the oven is operated at much lower temperature
\cite{pritchard93,ketterle05,straten07}.

Unlike the output of the Zeeman slower the output of the 2D~MOT source is
limited by a fundamental loss mechanism. As described in subsection
\ref{subsection:LossMechanisms} this is caused by Van der Waals forces between
atoms leaving the oven and (to a lesser extent) by resonant-dipole forces
between optically excited atoms in the 2D~MOT and the hot background flux from
the oven. These losses are quantified by the exponent in Eq.\thinspace
(\ref{eq:coldFluxFnT}), which is shown as the dashed line in Fig.\thinspace8.
Note that near maximum output at $T\approx700~\mathrm{K}$ the attenuation
factor is already as small as $\sim0.3$. Therefore, the source is best
operated at temperatures below $650~\mathrm{K}$, where the flux may be
slightly smaller but the depletion time of the oven is comfortably long.
Alternatively, one could incorporate a recycling principle \cite{hau94,hau05}.

\section{Summary and conclusion\label{section:Conclusion}}

We developed a novel beam source for cold $^{6}\mathrm{Li}$ atoms. The source
is based on the 2D~MOT principle and yields a controllable output flux of up
to $3\times10^{9}\,\mathrm{s^{-1}}$, comparable to fluxes typically achieved
in lithium Zeeman slowers. Some $30\%$ of the atoms are recaptured into a
3D~MOT $220~\mathrm{mm}$ downstream from the source. The source is side-loaded
from an oven and a push beam assures that only capturable atoms enter the main
vacuum chamber. This yields a clean and quite monochromatic cold atomic beam
of which the most-probable axial velocity $\alpha_{z}$ can be varied over the
range $18\lesssim\alpha_{z}\lesssim70~\mathrm{m/s}$ by varying the intensity
of the push beam. The 2D~MOT can be fully optimized for capture because the
push beam assures that the density of trapped atoms is intrinsically low. The
push beam also drastically simplifies the alignment of the 2D~MOT. Permanent
magnets simplify the implementation of the quadrupole field. The resulting
source is compact and enables us to load up to $10^{10}$ atoms into a 3D~MOT,
which is sufficient as a starting point for most experiments with quantum
gases. The output flux increases exponentially with the oven temperature until
at $T\approx700$~\textrm{K} a loss mechanism limits the flux. We identify
knock-out collisions near the oven exit as a result of Van der Waals forces
between the atoms as the limiting mechanism. At maximum output the beam
attenuation factor is $\sim0.35$. Therefore, the source is more efficiently
operated at a lower oven temperature. For $T=623$~\textrm{K} we measured a
loading rate of $\Phi_{r}=1.8(6)\times10^{8}\,\mathrm{s^{-1}}$ in the 3D~MOT.
At this temperature the uninterrupted running time on $8~\mathrm{g}$ of
lithium is $\sim17000$~\textrm{hrs}. With our work we demonstrate that the
2D~MOT principle works equally well with light atoms as with more heavy atoms
and is likely to be suitable for any atomic system with an optical cooling transition.

\section*{Acknowledgments}

The authors thank P. Cleary and M. Koot for assistance with the
characterization of the Nd$_{2}$Fe$_{14}$B magnets, G.V. Shlyapnikov and T.W.
Hijmans for stimulating discussions and N.J. van Druten for critically reading
the manuscript. This work is part of the research program on Quantum Gases of
the Stichting voor Fundamenteel Onderzoek der Materie (FOM), which is
financially supported by the Nederlandse Organisatie voor Wetenschappelijk
Onderzoek (NWO).


\end{document}